\documentclass[%
 aip,
pop,%
amsmath,amssymb,
reprint,%
]{revtex4-1}

\usepackage{xcolor}
\usepackage{graphicx}
\usepackage{dcolumn}
\usepackage{bm}

\newcommand{\nsample}{\ensuremath{n_\text{sample}}}

\renewcommand{\v}[1]{\ensuremath{\mathbf{#1}}} 
\renewcommand{\t}[1]{\ensuremath{\mathrm{#1}}}
\newcommand{\abs}[1]{\left| #1 \right|} 
\renewcommand{\d}[2]{\frac{d #1}{d #2}} 
\newcommand{\pd}[2]{\frac{\partial #1}{\partial #2}} 
\newcommand{\order}[1]{\ensuremath{\mathcal{O}\left( #1 \right)}}

\newcommand{\bvec}{\v{b}}
\newcommand{\kvec}{\v{\kappa}}

\newcommand{\Bvec}{\v{B}}

\newcommand{\jvec}{\v{j}}

\newcommand{\Ttilde}{\ensuremath{\tilde T}}

\newcommand{\qtilde}{\ensuremath{\tilde q}}

\renewcommand{\div}{\ensuremath{\nabla\cdot}}

\newcommand{\rbp}{\ensuremath{R\Bp}}

\newcommand{\bxgradgrad}[2]{\ensuremath{\bvec_0\times\nabla_\perp #1 \cdot\nabla #2 }}

\newcommand{\deltaEpar}{\ensuremath{\delta E_\parallel}}

\newcommand{\abar}{\ensuremath{\bar a}}
\newcommand{\thetabar}{\ensuremath{\bar\theta}}
\newcommand{\phibar}{\ensuremath{\bar\varphi}}
\newcommand{\xibar}{\ensuremath{\bar\xi}}

\newcommand{\modA}{\ensuremath{|A|}}

\renewcommand{\a}[2]{a_{#1#2}}

\begin{document}

\title{Machine Learning Surrogate Models for Landau Fluid Closure}

\author{Chenhao Ma}
\email{cma@pppl.gov}
\affiliation{Theory Department, Princeton Plasma Physics Laboratory, Princeton, NJ 08540}
 
\author{Ben Zhu}%
\email{zhu12@llnl.gov}
\affiliation{Fusion Energy Science Program, Lawrence Livermore National Laboratory, Livermore, CA 94550}%

\author{Xue-qiao Xu}%
\affiliation{Fusion Energy Science Program, Lawrence Livermore National Laboratory, Livermore, CA 94550}%

\author{Weixing Wang}
\affiliation{Theory Department, Princeton Plasma Physics Laboratory, Princeton, NJ 08540}

\date{\today}

\begin{abstract}
    The first result of applying the machine/deep learning technique to the fluid closure problem is presented in this paper. As a start, three different types of neural networks (multilayer perceptron (MLP), convolutional neural network (CNN) and two-layer discrete Fourier transform (DFT) network) were constructed and trained to learn the well-known Hammett-Perkins Landau fluid closure in configuration space. We find that in order to train a well-preformed network, a minimum size of the training data set is needed; MLP also requires a minimum number of neurons in the hidden layers that equals the degrees of freedom in Fourier space despite the fact that training data is fed in configuration space. Out of the three models, DFT performs the best for the clean data, most likely due to the existence of the simple Fourier expression for Hammett-Perkins closure, but it is the least robust with respect to input noise. Overall, with appropriate tuning and optimization, all three neural networks are able to accurately predict the Hammett-Perkins closure and reproduce the intrinsic \textit{nonlocal} feature, suggesting a promising path to calculating more sophisticated closures with the machine/deep learning technique.
\end{abstract}

\maketitle


\section{Introduction}
The fluid closure problem in plasma physics is probably as old as plasma physics itself. The problem arises when deriving fluid equations through the chains of moment equations from kinetic theories. The resulting lower order moment equations always contain a higher order moment. To truncate the moment hierarchy, a proper \textit{closure} is thus required to approximate this higher order moment from existing lower order moments for microscopic descriptions, which is conventionally constructed by phenomenological constitutive relations.
Moment closure hierarchies for kinetic theories are important and active scientific areas of research in fluid dynamics, plasma physics, neuroscience, and so on.
The widely used Spitzer-H\"arm closure~\cite{spitzer1953transport}, and similarly, Braginskii closure~\cite{braginskii1965transport} consider a strongly collisional plasma and predict heat flux $q\propto \nabla T$,
both of which lack kinetic effects and start to break down when the particle mean-free-path approaches the characteristic length scale (i.e., in weakly collisional regime).
Hammett and Perkins first proposed a so-called Landau-fluid closure as it incorporates Landau-damping effects in the electrostatic, collisionless, static limit\cite{hammett1990fluid}. Over the years, the Landau-fluid closure has been extended to collisional~\cite{chang1992unified,umansky2015modeling}, magnetized~\cite{guo2012parallel} plasma and dynamic perturbation conditions~\cite{hunana2018new,wang2019landau}. However, implementing Landau-fluid closures to high performance fluid codes is numerically challenging as the closures usually involve both frequency and wave-vectors in Fourier space~\cite{chang1992unified,hunana2018new,wang2019landau}. 

Over the same period of time, there has been significant progress in machine/deep learning method~\cite{spears2018deep}.
Inside the fusion plasma physics community, models have been trained to assist experimental data analysis~\cite{ferreira2018applications}, carry out real-time modeling of neutral beam injection~\cite{boyer2019real}, aid operation control (e.g., disruption prediction~\cite{cannas2007prediction,rea2018exploratory,kates2019predicting}), evaluate the core turbulent transport fluxes and the pedestal structure~\cite{meneghini2017self}, design better ICF experiments~\cite{peterson2017zonal} and so on. Here we are interested in another kind of machine learning application -- generating machine learning surrogate models of phenomena for microscopic descriptions that can be used within models for fluid simulations, yielding better and faster solutions. 
An interesting, open question is thus raised -- is it possible to get a machine-learned closure model for plasma physics? Or in other words, can a neural network be trained with existing theories, or data from simulations to learn the closure which could be further applied to fluid simulations? If the answer is affirmative, then how do we build and train such a model? How is its performance in terms of accuracy and speed-up?


As a proof of principle, we start with training neural networks to learn the Hammett-Perkins closure -- it is relatively simple (thus understandable), analytically solvable (thus easy to evaluate errors), yet non-trivial as it represents certain aspects of the general closure problems (e.g., the \textit{nonlocal} feature explained below). Furthermore, even though it is simple in Fourier space, it is difficult or impossible to implement in fluid simulation codes, such as BOUT++~\cite{dudson2009bout++} and GDB~\cite{zhu2018gdb}, in configuration space due to the particular discretizations, geometry and boundary conditions~\cite{dimits2014fast}. In such circumstances training the Hammett-Perkins closure in configuration space is desired.

\section{Physics model}
The Hammett-Perkins closure~\cite{hammett1990fluid} gives the relation between temperature and heat flux perturbation in Fourier space as
\begin{equation}
    \label{eq:closurek}
    \qtilde_k = -n_0\sqrt{\frac8\pi}v_t\frac{ik\Ttilde_k}{\abs{k}}.
\end{equation}
Transforming Eq.(\ref{eq:closurek}) from Fourier space back to configuration space yields
\begin{equation}
    \label{eq:closurez}
    \qtilde(x) = -n_0\sqrt{\frac8{\pi^3}}v_t\int_0^\infty dx'
    \left[\Ttilde(x+x') - \Ttilde(x-x')\right]/x'.
\end{equation}
The convolution in Eq.(\ref{eq:closurez}) indicates that the heat flux $\qtilde(x)$ is \textit{nonlocal} in configuration space, i.e.,
the perturbed heat flux $\qtilde(x)$ at location $x$ relies on not only the local temperature $\Ttilde(x)$ and its gradient $\partial\Ttilde(x)/\partial x$, but also the nearby temperature perturbations $\Ttilde(x\pm x')$.

The motivation of this study is to assess the neural networks' capability of recovering nonlocal feature. It shall be noted that in Fourier space the expression of Hammett-Perkins closure is linear and local, thus training a neural network for Eq.(\ref{eq:closurek}) becomes trivial. Therefore, in this paper, we focus on training neural networks to learn Hammett-Perkins closure Eq.(\ref{eq:closurez}) in configuration space. At the same time, we try to answer several technical questions such as: How much data is needed to train a neural network? Are there any constraints on neural network configuration? What type of neural network performs best and why? 

\begin{figure}
    \centering
    \includegraphics[width=\linewidth]{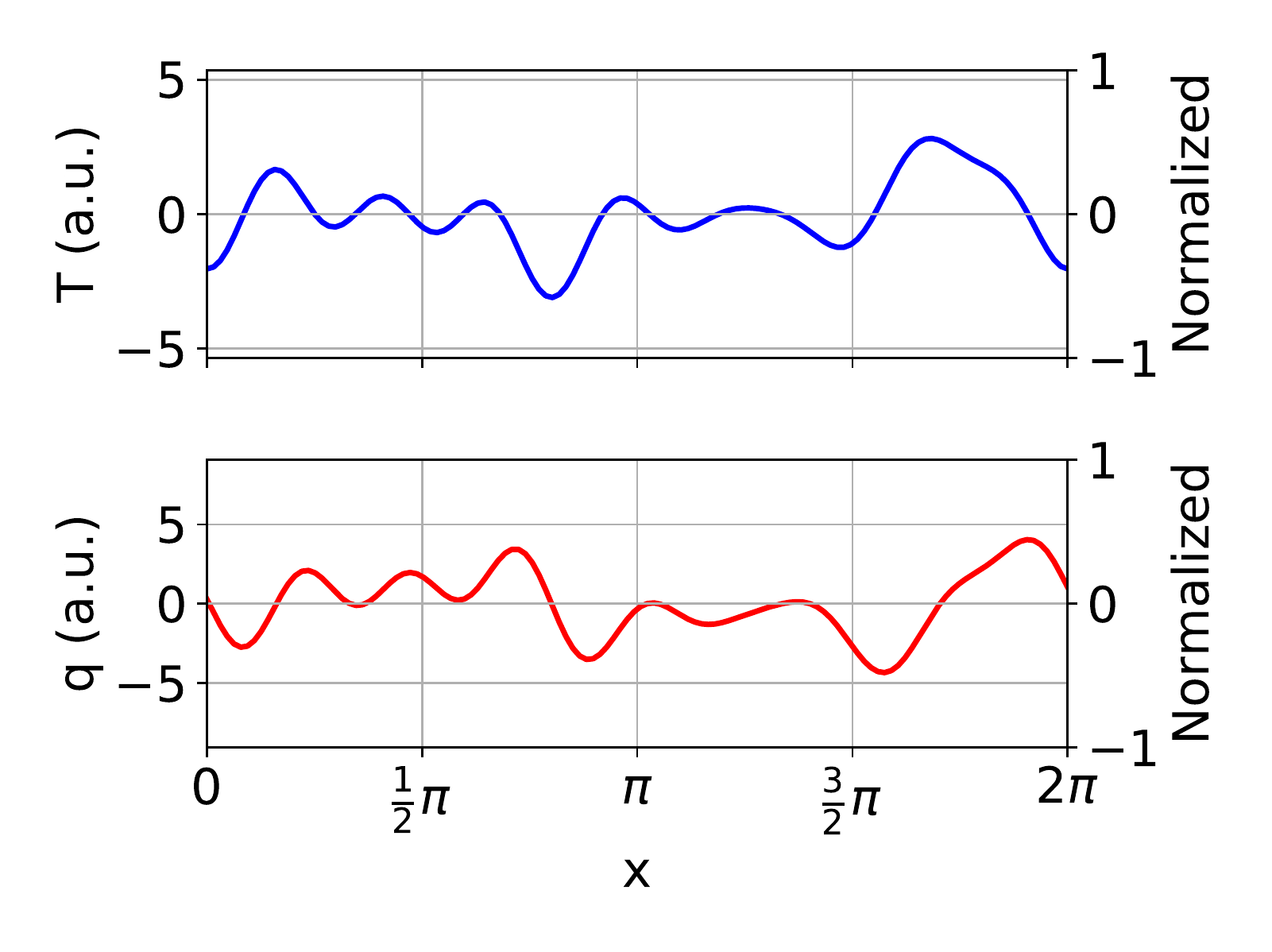}
    \caption{Fluctuating (a) temperature and (b) heat flux profiles.
    }
    \label{fig:profile}
\end{figure}

For simplicity, we assume that plasma has uniform density $n_0$, simulation domain $x\in[0,2\pi]$, resolution $n_x=128$ with periodic boundary condition. 
To prepare the data set for model training, validation and testing, a series of fluctuating temperature profiles $\{ \Ttilde_i(x) \}$ are first generated with
\begin{equation}
    \label{eq:data}
    \Ttilde_i(x) = \sum_{k=1}^{k_\text{max}}A_{ik}\sin(kx+\phi_{ik}),
\end{equation}
where uniformly distributed $A_{ik}$ and $\phi_{ik}$ are randomly chosen in the range $0\leq A_{ik}\leq1$ and $0\leq\phi_{ik}<2\pi$;
the corresponding $\qtilde_i(x)$ is then calculated with Eq. (\ref{eq:closurek}) by Fourier transform.
Fig. \ref{fig:profile} shows one pair of $\Ttilde(x)$ (blue) and $\qtilde(x)$ (red).
Finally, the paired temperature and heat flux profiles $\{ \Ttilde_i(x),\qtilde_i(x) \}$ are re-normalized to fit in between $-1$ to $1$,
\begin{equation}
    \hat{\Ttilde} = \frac{\Ttilde}{\Ttilde_{max}},\quad
    \hat{\qtilde} = \frac{\qtilde}{\qtilde_{max}},
\end{equation}
where $\Ttilde_{max}$ and $\qtilde_{max}$ are the maximum absolute values of $\Ttilde(x)$ and $\qtilde(x)$ in the entire data set, respectively.

\section{Machine learning surrogate models}
\begin{figure*}
    \centering
    \includegraphics[width=\linewidth]{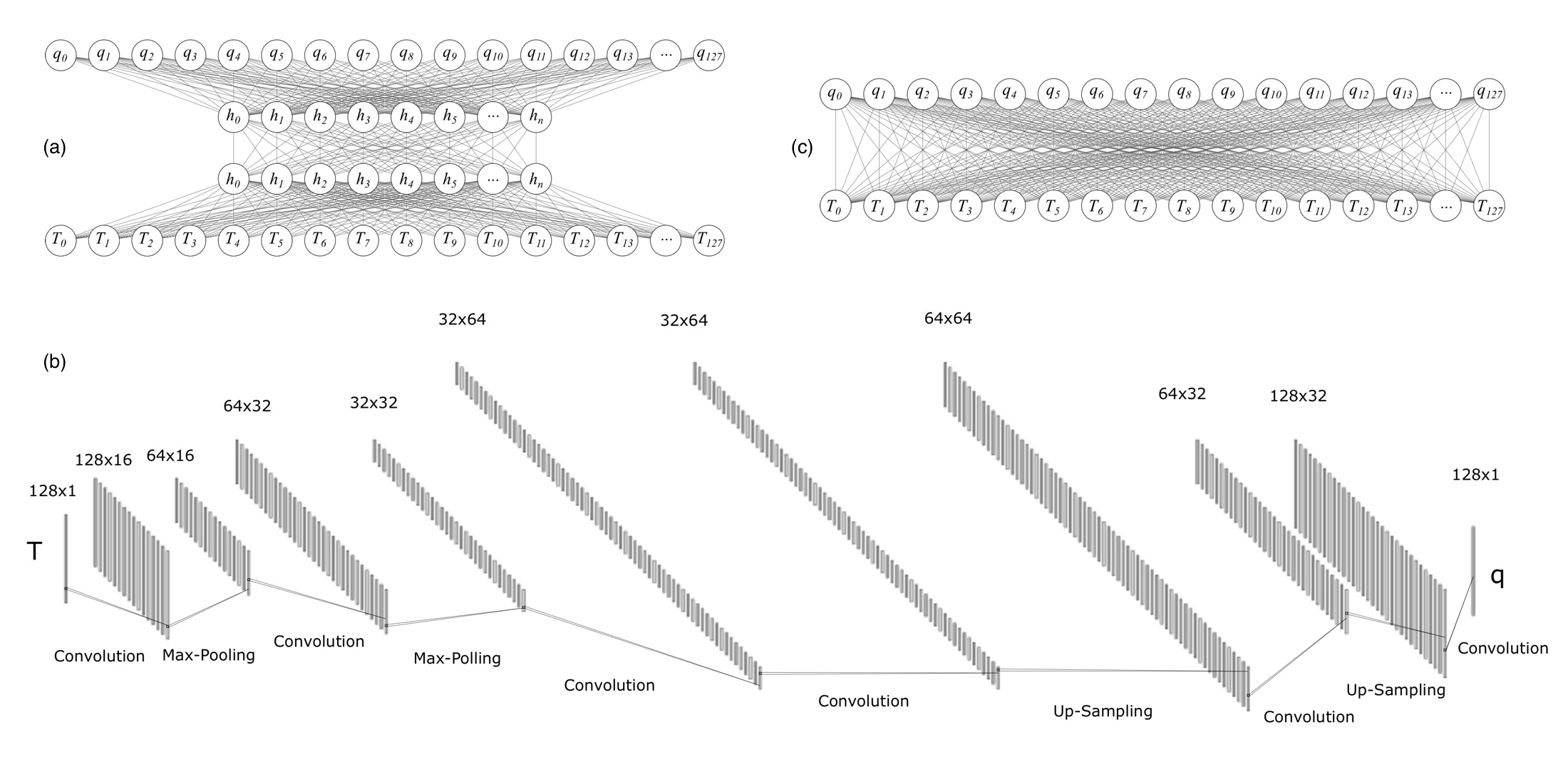}
    \caption{Neural Network architecture schematics of
    (a) multilayer perceptron (MLP): one input layer, two hidden layers and one output layer with 128 input and output variables and $n$ neurons for each hidden layer;
    (b) convolutional neural network (CNN): one input layer, double convolution-pooling layers, followed by two convolution layers, double upsampling-convolution layers, and one output layer with 128 input and output variables;
    (c) discrete Fourier transform neural network (DFT): one input layer and one output layer with 128 input and output variables and no hidden layers.
    }
    \label{fig:nn}
\end{figure*}

In this study, three types of neural networks are constructed with Keras interface~\cite{chollet2015keras}, 
and TensorFlow~\cite{tensorflow2015-whitepaper} backend as shown in Figure~\ref{fig:nn}.
The activation and loss functions are two key concepts in neural networks. The former defines the output of a neuron given an input or a set of inputs. Rectified linear unit (ReLU) is the most commonly used activation function in deep learning, which is defined as $f(x) = \max\{x, 0\}$.
The latter defines a mean to evaluate model performance by comparing the discrepancy between the estimated and true values of the solution. Two popular loss functions for the regression problem are mean-squared-error (MSE) and mean-absolute-error (MAE), which are defined as $\epsilon_{mse}=\sum_i|\hat{y}_i-y_i|^2/N$ and
$\epsilon_{mae}=\sum_i|\hat{y}_i-y_i|/N$. Here $\hat{y}_i$ and $y_i$ are the estimated and true values, respectively, and $N$ is the number of samples.

The first type of neural network we train is a conventional multilayer perceptron (MLP)~\cite{haykin2009neural}.
It contains one input layer, two hidden layers and one output layer.
Either ReLU or the tanh function is chosen as the activation function in MLP except on the last layer where linear function is applied.
The second type of neural network is a convolutional neural network (CNN)~\cite{lecun2015deep}.
It has an input layer, double convolution-pooling layers, followed by two convolution layers, and double upsampling-convolution layers.
In CNN, the activation function is ReLU for hidden layers, 
and linear for the output layer.
The third type of neural network is a two-layer discrete Fourier transform (DFT) neural network~\cite{velik2008discrete}. 
All three networks use standard Adam optimization algorithm~\cite{kingma2014adam}, and MSE as the loss function for training and validation. However, in the following discussion, we primarily use MAE and the mean of MAE distribution $\bar{\epsilon}_{mae}$ as the error metric since it is more intuitive, i.e., no square-root operation is needed to evaluate the relative error.

\section{Training results}

Neural networks are data-hungry, i.e., the performance of the neural networks strongly depends on the amount of training data available. A well trained neural network shall also be neither overfit nor underfit~\cite{spears2018deep}, i.e., it should have a minimum discrepancy between training and validation loss. The first task is therefore trying to find out how much data is adequate for this simple 1D closure problem by comparing the training loss and the validation loss.
We start to train the MLP model with an increasing size of training data $\nsample$.

\begin{figure}
    \centering
    \includegraphics[width=\linewidth]{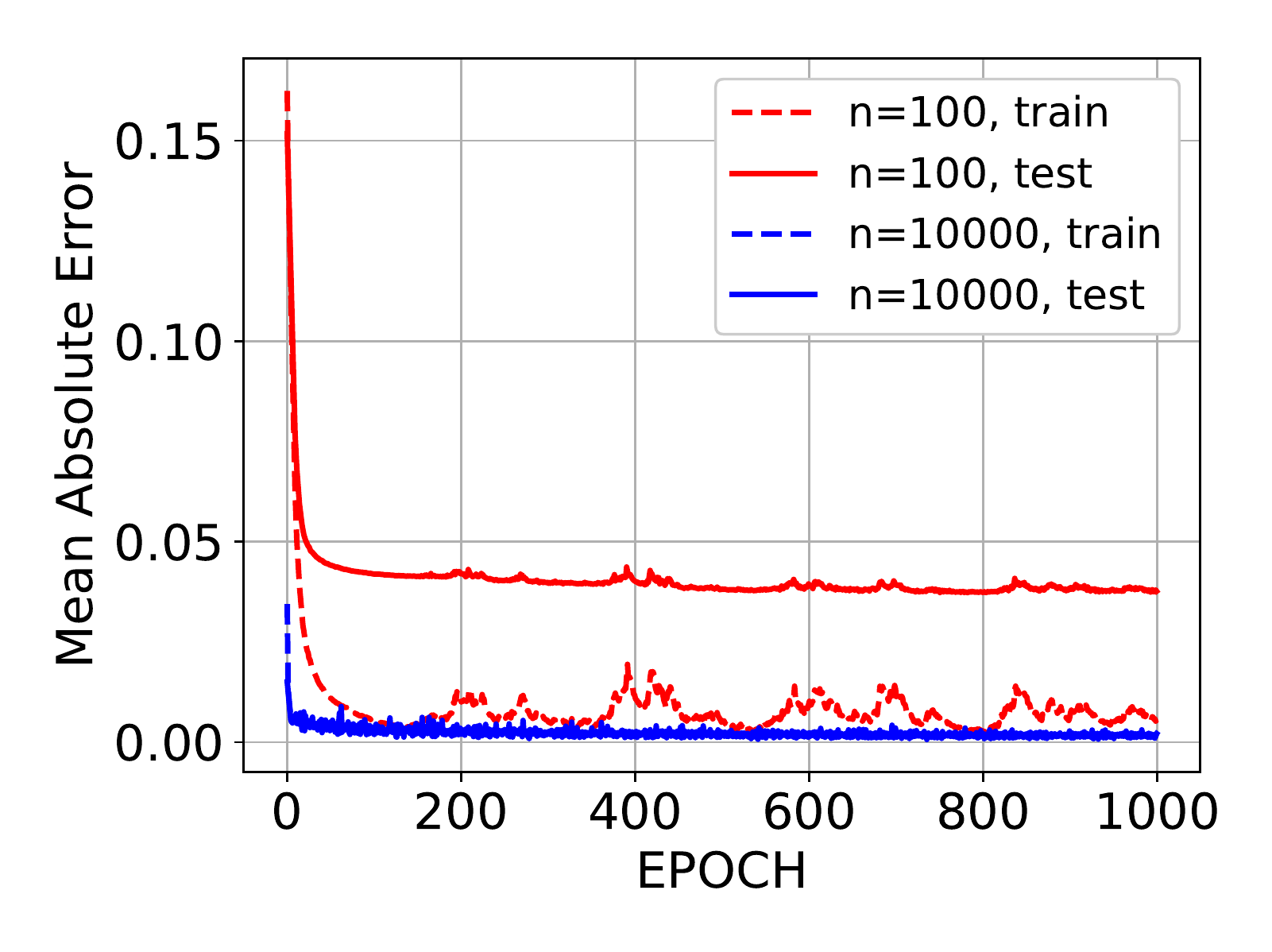}
    \caption{Training history of MLP with different data sizes: (red) $\nsample=100$ and (blue) $\nsample=10,000$.
    Dashed and solid lines denote training and testing Mean-Absolute-Error respectively.}
    \label{fig:maehistory}
\end{figure}

Fig. \ref{fig:maehistory} illustrates the training history of MLP with $\nsample=100$ and $\nsample=10,000$. 
Here the ``epoch" in the horizontal axis means the number of times that the entire training data set is passed to the neural network.
In both cases, training and testing MAEs start at a similar value.
During the training process, the training MAE decreases much faster than the testing MAE for $\nsample=100$ case; and the discrepancy between training and testing MAEs at the converged phase indicates that the MLP is overfitted, suggesting the training data set is insufficient.
For the $\nsample=10,000$ case,
the training MAE and the testing MAE have the similar values almost over the whole training process.
We hence conclude that at least $\nsample=10,000$ training data is required for the 1D closure problem at this resolution for MLP.
Fig. \ref{fig:maehistory} also shows that the trained model tends to converge faster with a larger data set: when $\nsample=100$, model converges near epoch=150; while for $\nsample=10,000$, it converges near epoch=60. 


\begin{figure}
    \centering
    \includegraphics[width=\linewidth]{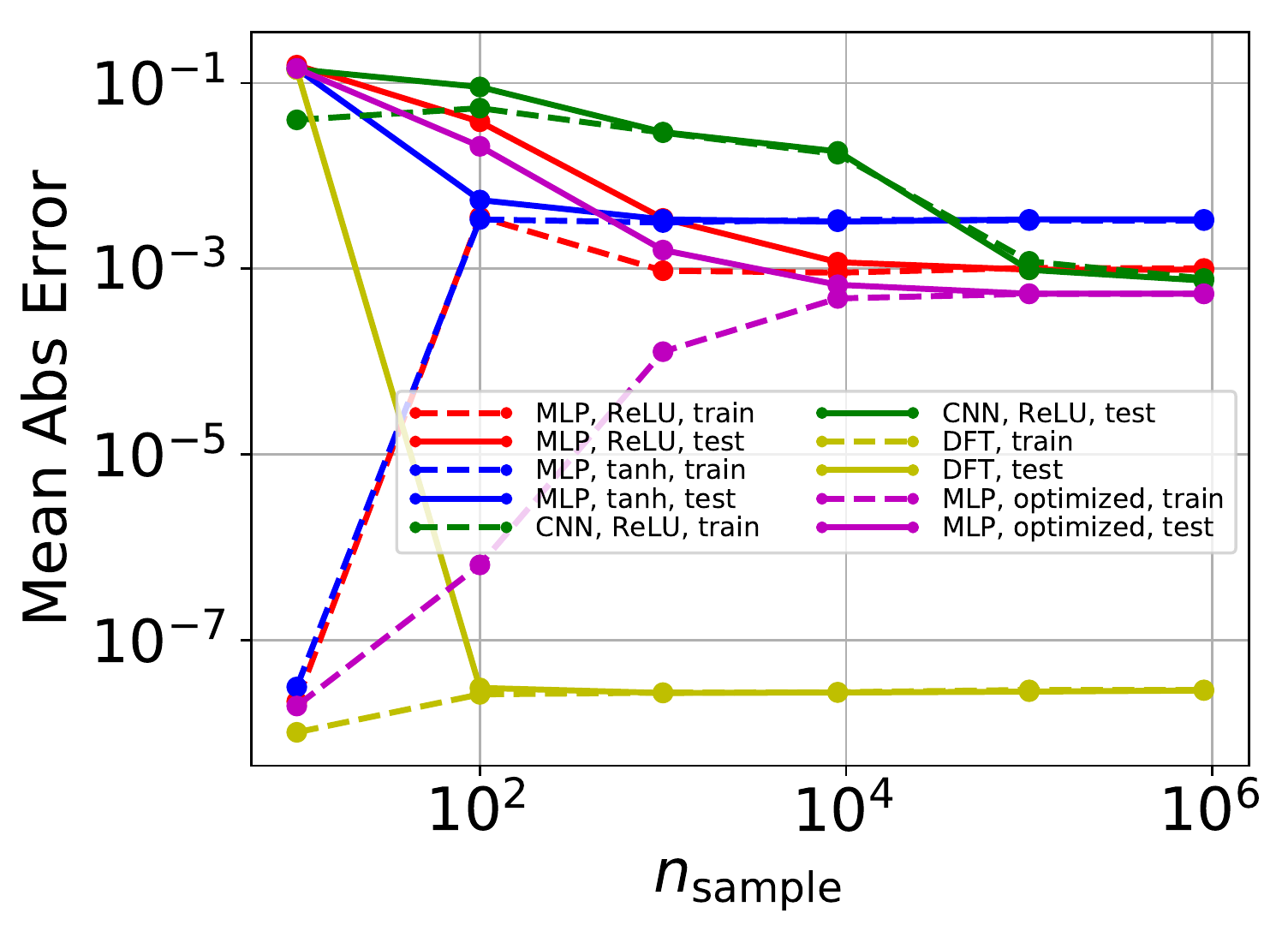}
    \caption{
    Mean-Absolute-Error with the number of training samples. 
    Dashed and solid lines denote training and testing MAE respectively;
    red, blue and green lines represent MLP with ReLU, MLP with tanh and CNN with ReLU as activation function respectively.
    Yellow line represents DFT result.
    Purple line represents the result of Bayesian optimized model.
    }
    \label{fig:maensample}
\end{figure}

In machine/deep learning, ``hyperparameter" refers to a parameter that is chosen before the training process, e.g., the activation function selection, hidden layer configuration, optimization algorithms and so on. Choosing the correct hyperparameter is critical for model performance. Here, we explore the impact of different activation functions.
Fig. \ref{fig:maensample} shows the MAE of different types of neural network 
with different activation functions and different $\nsample$ used in the training.
In general, the discrepancy between the training MAE and the testing MAE decreases when $n_\text{sample}$ increases from $10$ to $1,000,000$, except for MLP with $\nsample=10$ when the training MAE is the 6th order of magnitude smaller than testing MAE.
As discussed above, the apparent departure is an indicator that the model is either overfit or underfit.
In this case, the MLP is clearly overfit as \nsample is too small and the resulting model poorly predicts the heat flux for testing data.
On the other hand, the near-perfect overfit with $\nsample=10$ result proves that fully connected MLP is indeed able to capture the nonlocal effect of Hammett-Perkins closure Eq.(\ref{eq:closurez}).
When \nsample~increases, the training MAE increases since model can no longer overfit all the training data. Meanwhile, the testing MAE decreases, until they converge to the same value when \nsample~is large enough.
Test results also show that the MLP with tanh function converges with less data at $\nsample=100$,
and gives a smaller testing MAE than the MLP with ReLU function when $\nsample\lesssim 1,000$.
This is because the tanh function suffers from the vanishing gradient problem
which prevents the neural network from further training even with more data.
Thus, MLP with ReLU has better performance when training data is larger (e.g., $\nsample>10,000$ in our test).
The purple line shows the results of the neural network which utilizes the Bayesian approach to optimize the number of layers and the number of neurons per layer for MLP, the most accurate MLP model is found to be one hidden layer with 47 neurons, and the resulting $\epsilon_{MAE}^{\text{Bayesian}} = 5.49\times10^{-4}$ after 100 optimization iterations. This is a $43\%$ accuracy improvement compared to two layer model (red line).
Yellow lines shows the result of DFT network, which has a testing MAE $\epsilon_{DFT}=2.69\times10^{-8}$.
This is because it represents a linear matrix multiplication.
The DFT network will be discussed later.

\begin{figure}
    \centering
    \includegraphics[width=\linewidth]{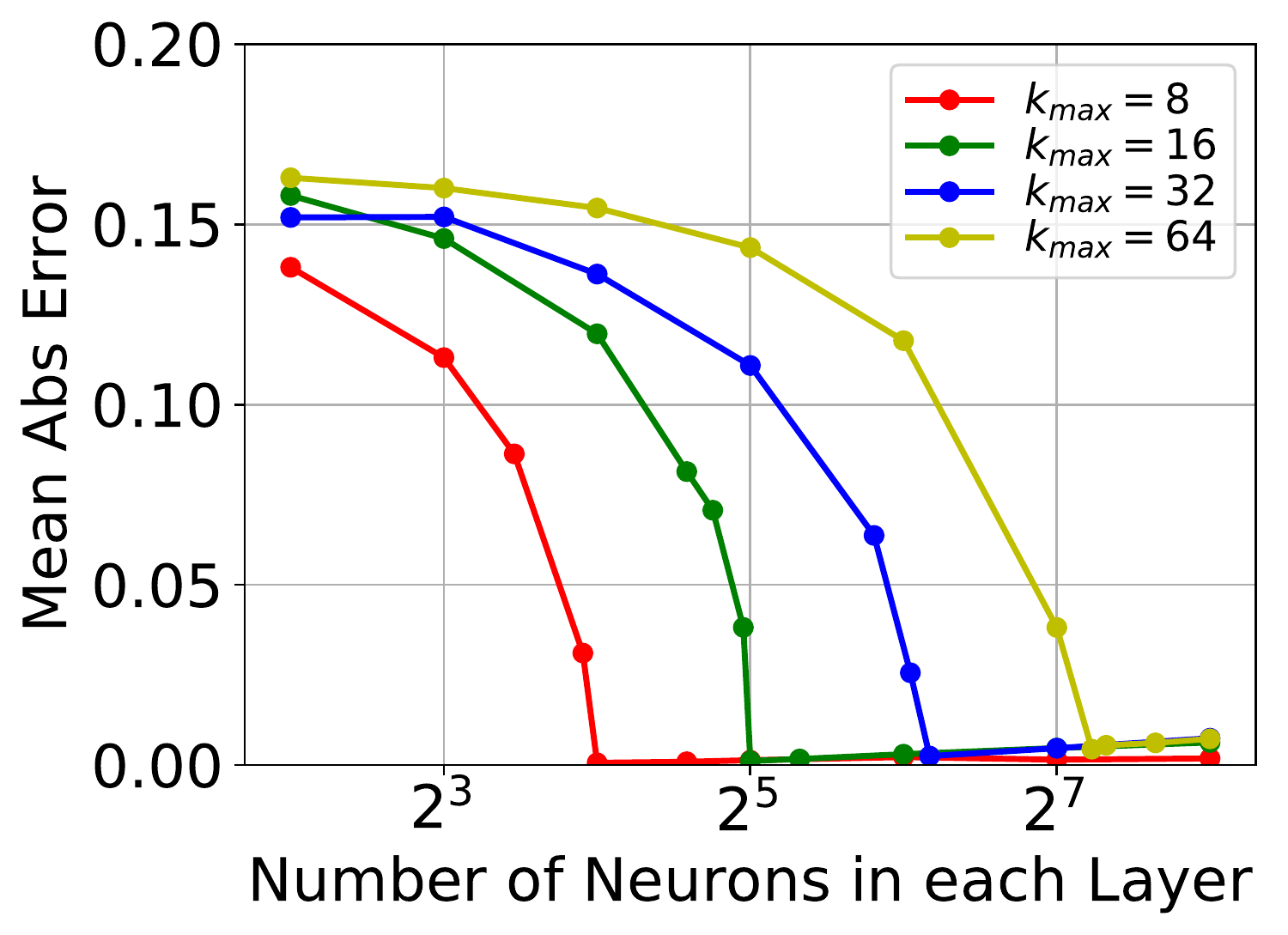}
    \caption{Mean-Absolute-Error of MLP for different numbers of neurons in each layer.
    red for $k_\text{max}=8$,
    green for $k_\text{max}=16$,
    blue for $k_\text{max}=32$,
    yellow for $k_\text{max}=64$.
    }
    \label{fig:maetwolayer}
\end{figure}

The width and depth of neural networks (i.e., the number of neurons in the hidden layers $n_\text{neuron}$ and the number of hidden layers) are also crucial hyperparameters in a neural network.
Here we first test the impact of $n_\text{neuron}$ on the testing MAE for a two hidden layers MLP.
Fig. \ref{fig:maetwolayer} shows that the testing MAE varies with $n_\text{neuron}$ for different numbers of modes $k_\text{max}$ in the system.
Because the resolution is set to be $n_x=128$ for $\Ttilde$ and $\qtilde$,
the maximum number of allowable Fourier modes in the system is 64.
We scan the $k_\text{max}$ in Eq.(\ref{eq:data}) from 8 to 64 by generating separated data sets,
and training the neural network with $\nsample=10,000$.
The sudden dropping of MAE for all cases demonstrates that there is a threshold of minimum number of neurons required in each hidden layer $n_\text{neuron}$ in order to accurately recover Hammett-Perkins closure.
The threshold is approximately
\begin{equation}
    n_\text{neuron}^\text{th} = 2k_\text{max}.
\end{equation}
When $n_\text{neuron}< n_\text{neuron}^\text{th}$, the testing MAE is always larger than 0.01; while $n_\text{neuron}\geq n_\text{neuron}^\text{th}$ the testing MAE rapidly reduces an order of magnitude or more.
The threshold $2k_\text{max}$ coincides with the degrees of freedom of the input $\Ttilde(x)$ in Fourier space representation -- that is, amplitude $A_k$ and phase shift $\phi_k$ are randomly chosen for each $k$, suggesting MLP is able to learn the Fourier space information despite the training data is fed in configuration space.
This also implies that if the input data contains full-spectrum information, then $n_\text{neuron}$ shall be at least equal to $n_x$ in order to resolve all Fourier modes.
It should also be noted that based on our tests with fixed number of training samples $\nsample=10,000$, increasing $n_\text{neuron}$ well beyond $n_\text{neuron}^\text{th}$ does not necessary improve model performance dramatically especially for the low $k_\text{max}$ cases.
As for the depth of a model, no significant improvement in MAE was found as the number of hidden layers was gradually increasing from 1 to 6 ($\epsilon=8.89\times 10^{-4}$ to $\epsilon=1.15\times 10^{-3}$). However, reducing the number of layers may reduce computational cost.
Therefore, we conclude (1) the minimum number of neurons of hidden layers depends on the input training data -- in general, $n_\text{neuron}$ should be a little bit larger than $n_\text{neuron}^\text{th}=2k_\text{max}$ where $k_\text{max}$ is the maximum wave number in the input data, and (2) a two-hidden-layer MLP is sufficient to accurately calculate the heat flux $\qtilde_i(x)$ for $\Ttilde_i(x)$ given $n_\text{neuron}\geq n_\text{neuron}^\text{th}$.
Unless noted otherwise, results in the rest of the paper were obtained with training data set with $k_\text{max}=8$ and if applicable, a two-hidden-layer MLP with $n_\text{neuron}=64$.

We also train one classical deep learning model -- the convolutional neural network (CNN). Here we pick a widely used CNN architecture in computer vision studies. It consists of an input and an output layer, as well as multiple convolution layers, pooling layers and up-sampling layers as hidden layers.
The convolution layers and pooling layers generate a latent space representation for the input temperature profile,
and the convolution layers and up-sampling layers decode the latent space representation to the output flux profile.
The output of convolution layer is the sliding dot product of its input and its \textit{filters} where the \textit{filters} detect the specific type of local features of the input.
The weights of a filter are shared in the whole domain, which is reasonable assuming spatial symmetry.
Pooling layer down-samples the input by combining multiple neurons in the previous layer into one neuron in the next layer,
while up-sampling layer duplicates one neuron in the previous layer to multiple neurons in the next layer.
As shown in Fig.~\ref{fig:maensample}, CNN appears unable to perfectly overfit the training data with $\nsample=10$, but eventually has a similar performance as MLP when the training and testing MAE converged at $\nsample=100,000$.
This shows that CNN is more data hungry because it estimates multiple learnable parameters and requires more sample data to converge.

\begin{figure}
    \centering
    \includegraphics[width=\linewidth]{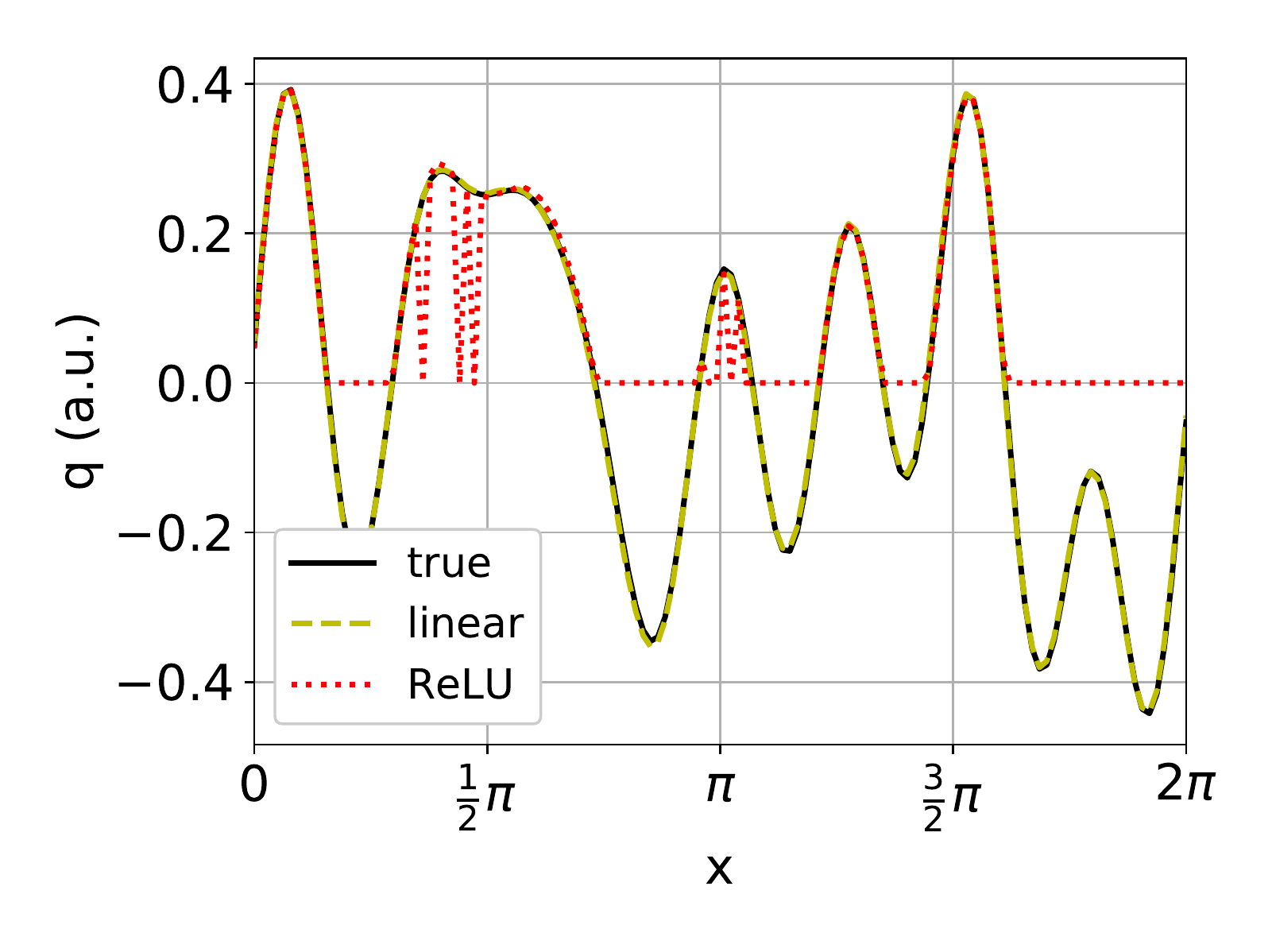}
    \caption{Effect of activation function of output layer.
    The black solid line represents the true value;
    the yellow dashed line represents the prediction with linear function;
    the red dotted line represents the prediction with ReLU function.
    }
    \label{fig:maeactivation}
\end{figure}

Due to the nature of the closure problem, the choice of activation function of the output layer is essential to getting correctly trained MLP and CNN.
Because the heat flux profile has both positive and negative values, 
ReLU is no longer suitable as the activation function of the last layer since it only outputs in $[0,\infty)$. 
Fig. \ref{fig:maeactivation} shows an example that the CNN fails to predict the output $\qtilde$
when ReLU is chosen as the activation function of the output layer.
An appropriate activation function of the last layer is linear function. When the activation function is switched from ReLU to linear, the CNN correctly predicts the output $\qtilde$ from the input $\Ttilde$.

The last type of neural network we train is a two-layer discrete Fourier transform (DFT) neural network. DFT is different than MLP and CNN in the sense that it is intentionally designed to solve Fourier transform type problem.
It has no hidden layer -- only the input and output layers; the activation function is linear for all neurons; and it trains transformation matrix $\mathcal{W}$. 
For instance, in DFT the relation between input $\Ttilde(x)$ and output $\qtilde(x)$ is $\qtilde=\mathcal{W}\Ttilde$.
Recall the definition of forward and inverse discrete Fourier transform
\begin{equation}
    \label{eq:dft}
    X_k = \mathcal{F}x_n, \quad x_n = \mathcal{F}^{-1}X_k, 
\end{equation}
where $x=\{x_n\}$ and $X=\{X_k\}$ are the signals in configuration and Fourier space, $\mathcal{F}$ and $\mathcal{F}^{-1}$ are forward and inverse Fourier transfer matrix.
The idea behind DFT neural network is training weight matrix $\mathcal{W}$ to be close to the analytical form of forward transform matrix $\mathcal{F}=\{\mathcal{F}_{jk}\}=\{\omega^{jk}/\sqrt{N}\}$ or inverse transform matrix $\mathcal{F}^{-1}=\{\mathcal{F}_{jk}^{-1}\}$ where $N$ is the total sample in one period (i.e., $n_x$ in our setup).
Back to Hammett-Perkins closure Eq.(\ref{eq:closurez}), it can be rewritten in DFT matrix form
\begin{equation}
    \label{eq:dft_hp}
    \qtilde(x) = \mathcal{F}^{-1}\mathcal{K}\mathcal{F}\Ttilde(x),
\end{equation}
where $\mathcal{K}=\{k_{jk}\}=\{-i\text{sign}(k)\delta_{jk}\}$ is the matrix form of $-ik/\abs{k}$.
Therefore training a DFT neural network to learn Hammett-Perkins closure in configuration space is equivalent to learning weight matrix $\mathcal{W} = \mathcal{F}^{-1}\mathcal{K}\mathcal{F}$. 
One would hence expect that DFT shall have a good performance due to the existence of analytical expression (Eq.(~\ref{eq:dft_hp})) for Hammett-Perkins closure.

\begin{figure*}
    \centering
    \includegraphics[width=\linewidth]{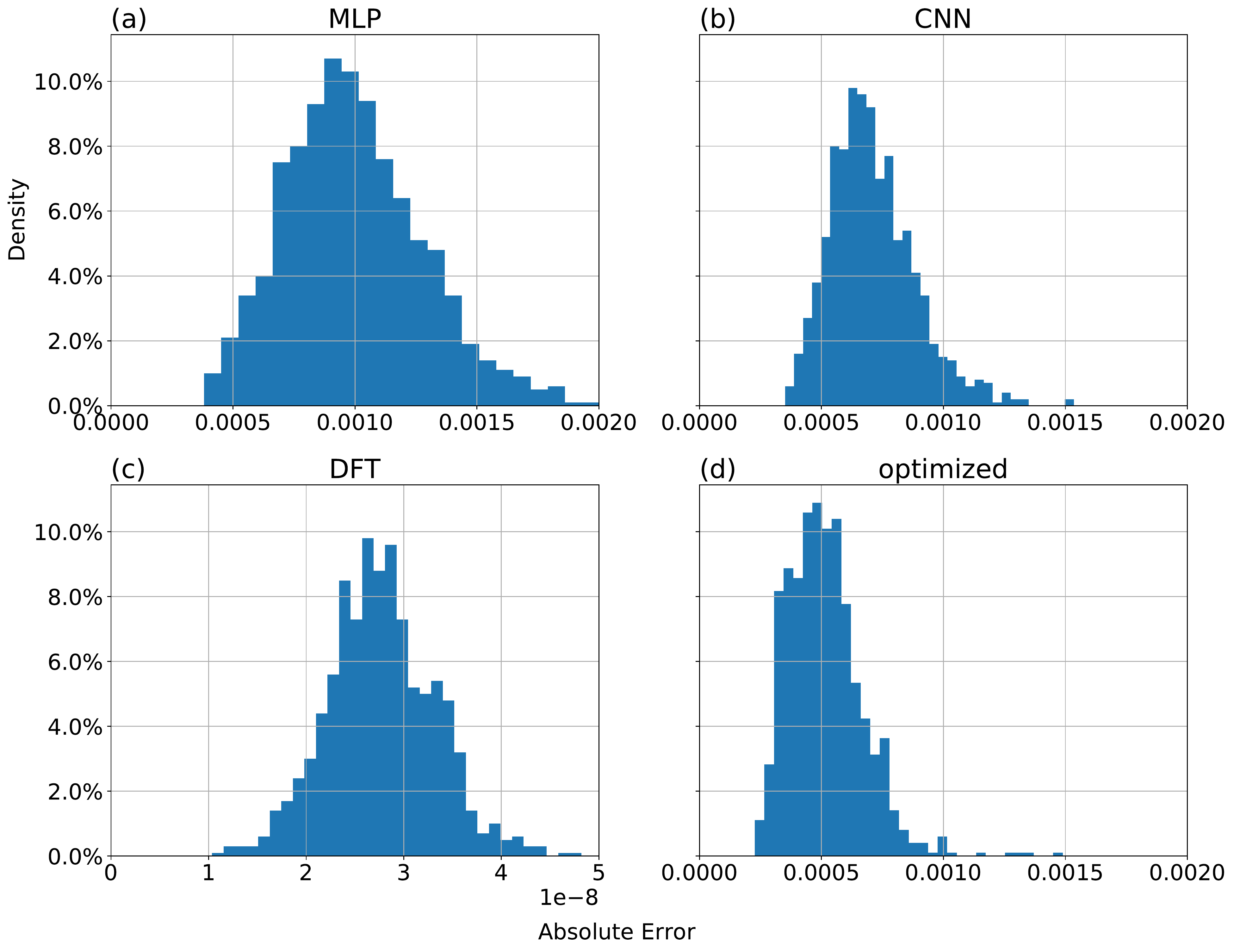}
    \caption{Probability distribution function (PDF) of absolute error $\epsilon$ for different neural network models when $\nsample=10^6$.
    (a) for MLP, (b) for CNN, (c) for DFT, and (d) for Bayesian optimized model.
    The mean absolute errors for different NN are
    $\epsilon_{MLP}=9.72\times10^{-4}$,
    $\epsilon_{CNN}=7.45\times10^{-4}$,
    $\epsilon_{DFT}=2.69\times10^{-8}$.
    }
    \label{fig:maehist}
\end{figure*}

The accuracy of the three different types of neural network, indicated by the probability distribution function of the mean-absolute-error (MAE), is summarized in Fig. \ref{fig:maehist}.
All three models are trained with the same training data set with $\nsample^\text{train}=1,000,000$, and tested with the same test data set with $\nsample^\text{test}=10^4$. Note that the test data set is independent of training data set.
In this test, both MLP and CNN use ReLU as the activation functions for the hidden layers.
The original two hidden layers MLP and CNN turn out to have similar accuracy with CNN being slightly better than MLP ($\epsilon_{MLP}=9.72\times10^{-4}\gtrsim \epsilon_{CNN}=7.45\times10^{-4}$); while DFT is 4 orders more accurate than MLP and CNN ($\epsilon_{DFT}=2.69\times10^{-8}$).
The optimized MLP model by the Bayesian approach gives a more accurate prediction with $\epsilon_{MAE}^{\text{Bayesian}} = 5.49\times10^{-4}$ and its error distribution has a significant narrower ($\sim 50\%$) width compared with the original MLP error distribution.
The impact of $10^{-3}$ level error from the neural network on its applications, e.g., whether error accumulates in a long-term fluid simulation, is unclear at this point and is subject of further studies.

\begin{figure}
    \centering
    \includegraphics[width=\linewidth]{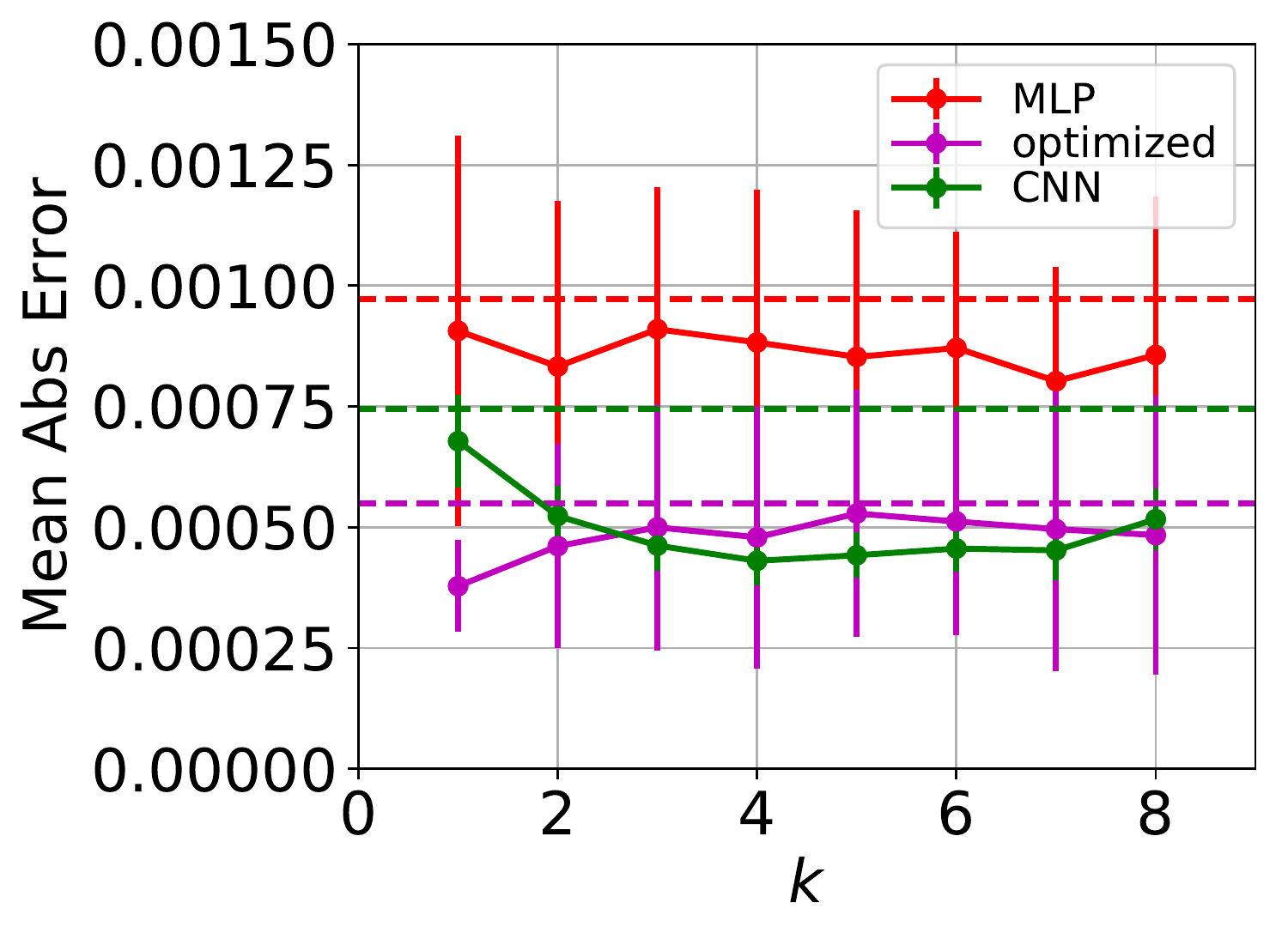}
    \caption{The MAE(dots) and STD(bars) of the prediction errors of MLP (red), CNN (green), and Bayesian optimized MLP (purple) if the testing data only contains single $k$ perturbation.
    }
    \label{fig:singlek}
\end{figure}

Although both MLP and CNN are trained with mixed multi-mode data, they are able to resolve individual mode as well.
Fig. \ref{fig:singlek} shows how the MAE and standard deviation (STD) of the prediction errors change for single $k$ testing data.
The prediction errors of MLP (red line), CNN (green line) and optimized MLP (purple line) do not change significantly with the mode number $k$, suggesting all modes in the system are resolved equally well.
It should be note that CNN appears to be able to handle single mode cases exceptionally well -- for MLP, the STD of single mode data has the same order as the STD of multiple mode data;
for CNN, however, the STD of single mode data is an order smaller than the STD of multiple mode data.
In multiple mode cases, MLP and CNN have similar deviations in the prediction error (Fig.~\ref{fig:maehist}).

\begin{figure}
    \centering
    \includegraphics[width=\linewidth]{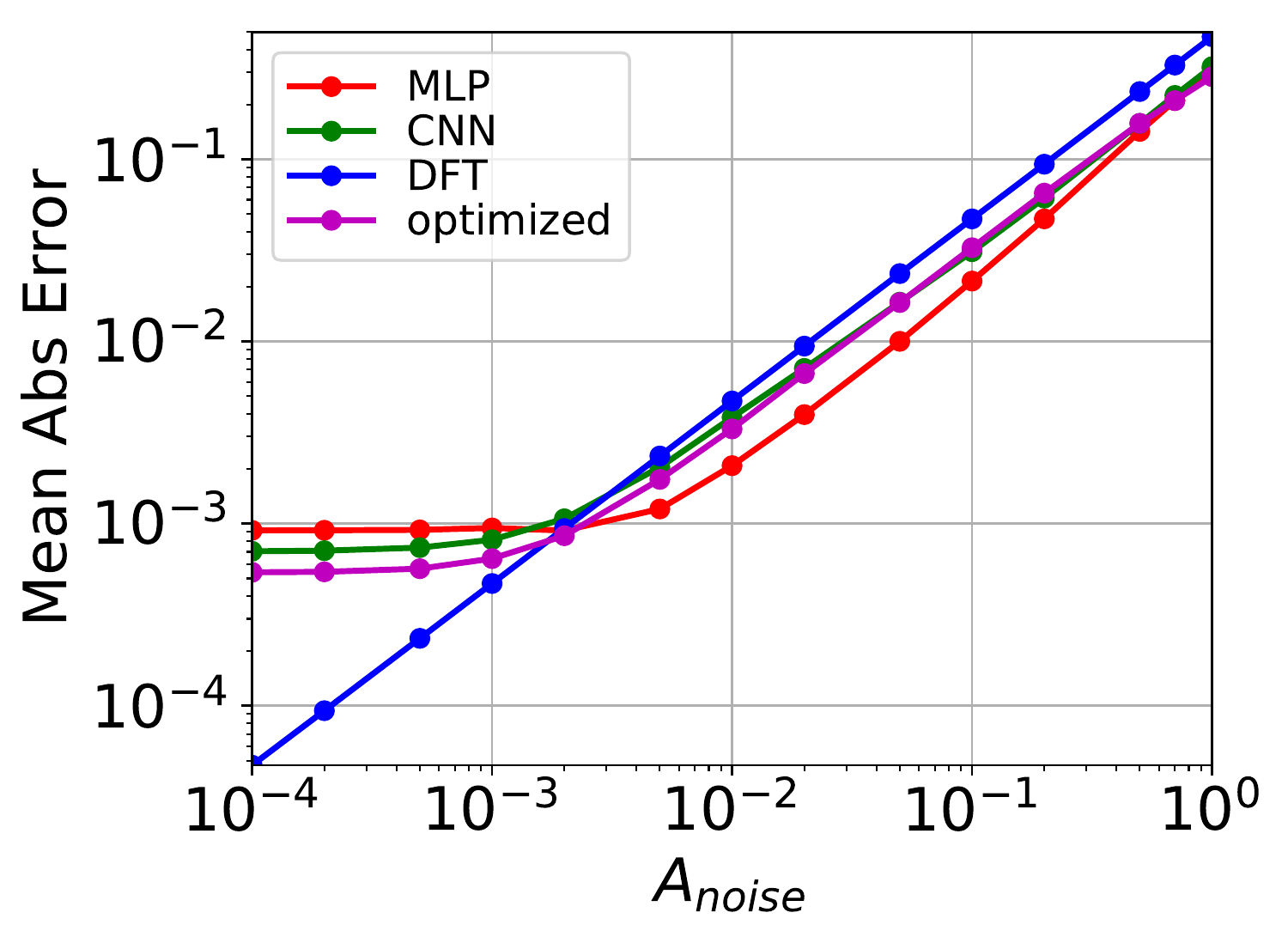}
    \caption{The MAE with the amplitude of noise.
    red line for MLP,
    green line for CNN,
    blue line for DFT,
    and purple line for Bayesian optimized model.
    }
    \label{fig:noise}
\end{figure}

A successful machine learning model should also be robust with respect to the uncertainty (or noise) presented in simulation and experimental data due to truncation errors, random/stochastic processes, and/or the precision and accuracy of the measuring instrument. We therefore study the robustness of trained models by adding white noise $A_\text{noise}\Ttilde_\text{noise}$ to the test data set $\Ttilde$ and monitoring the model prediction $\qtilde_{\text{signal}+\text{noise}}$. Here $A_\text{noise}$ is the maximum amplitude of white noise, and the error metric MAE is evaluated by comparing $\qtilde_{\text{signal}+\text{noise}}$ to the true $\qtilde$ with clean $\Ttilde$.
Fig. \ref{fig:noise} shows how MAE varies with $A_\text{noise}$ for three different models.
The change of MAE for MLP and CNN has two phases:
(1) when $A_\text{noise}$ is smaller than the prediction error of neural network (roughly $\epsilon$), the MAE does not increase with $A_\text{noise}$ as $\epsilon$ over-dominates $A_\text{noise}$;
(2) when $A_\text{noise}$ is larger than $\epsilon$, the MAE starts to increase linearly with the noise amplitude $A_\text{noise}$.
It also appears that MLP with two hidden layers is more resilient to noise, most likely because the trained network filters out the high-$k$ noise with its multiple layers.
We remark that although Bayesian optimized model performs better than the two hidden layers MLP for the clean data, it gives less accurate predictions for the noisy data when $A_\text{noise}>2\times10^{-3}$.
In contrast, DFT seems to capture all levels of white noise because DFT essentially mimics the analytic expression of Hammett-Perkins closure and cannot ``filter" any noise.
It thus becomes less accurate than MLP or CNN when $A_\text{noise}>2\times10^{-3}\sim \epsilon_{MLP,CNN}$.
Overall, MLP is the most robust model and DFT is the least robust model in the noise test.
The noise level from PIC simulation is in the order of $1/\sqrt{n}$, where $n$ is the number of particles per cell. Typically $n=50$ in nonlinear PIC simulation; therefore, the noise level is around 14\%. The two-hidden-layer MLP is the most accurate model among the ML surrogate model we studied in the paper.

\section{Summary and Outlook}

Even though MLP and CNN are similar in performance for the 1D closure problem, we suspect that CNN will perform much better than MLP for the future 2D and 3D closure problems as the convolution layer is even more efficient at handling 2D and 3D data~\cite{lecun2015deep}.
In principle, the fully connected MLP can also be used to learn the features from 2D and 3D data.
However it becomes less practical as the number of neurons and weights are increasing greatly with data dimension.
For example, the number of weights for 3D data is $n_xn_yn_z$ for each neuron in the a fully connected layer.
On the other hand, the number of weights of a filter in the convolution layer is $n_{f,x}n_{f,y}n_{f,z}$ where $n_{f,i}<n_i(i=x,y,z)$ is the filter size in each direction, independent of the input size.
In our current 1D study with 128 inputs/outputs, the evaluation cost for MLP is $11\mu s$ per sample, for DFT it is $10\mu s$ per sample, and for CNN it is $115\mu s$ per sample. Although the evaluation cost for CNN is around an order of magnitude larger than MLP and DFT for this 1D case, it scales quadratically for MLP and DFT as $O(n^2)$ while linearly for CNN scales as $O(n_kn)$, where $n$ is the problem size and $n_k=15$ is the kernel size. Therefore, for a 2D problem with $128\times128$ inputs/outputs, evaluation costs for MLP, DFT and CNN will approximately be 180$ms$, 164$ms$ and 14$ms$ respectively.
It is not feasible to use MLP or DFT in large scale 2D and 3D problems.

Generally the fully connected layer performs better than the convolution layer for 1D problem, such as the 1D Hammett-Perkins closure we discuss here.
As for DFT, despite the fact that it has superior accuracy compared to the MLP and CNN, it may lose its advantage when training advanced closures with complicated, if not impossible, analytical forms, e.g., data from the fully nonlinear simulation of Vlasov-Poisson equations. 
The performance of MLP, CNN and DFT for 2D/3D closure problems is beyond the scope of this study and is left to be explored in the future.

We emphasize that our work here is a proof-of-principle study to demonstrate that appropriate neural networks are capable of calculating Landau fluid closure.
It is our hope that the success of training machine-learned Hammett-Perkins closure would help researchers to gain confidence in the machine/deep learning approach of closure problem and also guide our future attempts on training more sophisticated closure models.
Implementation of such a neural network model in global 3D fluid turbulence codes, for example, BOUT++~\cite{dudson2009bout++} and GDB~\cite{zhu2018gdb}, is on another level of complexity (e.g., nonuniform grid, non-periodic boundary conditions, dependence on other plasma quantities). Further benchmarking neural network based Landau fluid code with existing Landau-fluid~\cite{ma2015impact} and Gyro-Landau-fluid module~\cite{ma2015KBM,ma2016global} of BOUT++ is an on-going research and will be reported in future publications.

\section*{Acknowledgments}
The authors would like to thank Drs. X.Z.Tang, R.Maulik, C.K.Huang, L.Wang, C.F.Dong for the helpful discussions.
C.H.Ma and W.X.Wang are supported by DOE contract DE-AC02-09CH11466 for the Princeton Plasma Physics Laboratory. 
B.Zhu and X.Q.Xu are supported by DOE contract DE-AC52-07NA27344 for the Lawrence Livermore National Laboratory. 
This research used resources of the National Energy Research Scientific Computing Center, a DOE Office of Science User Facility supported by the Office of Science of the U.S. Department of Energy under Contract No. DE-AC02-05CH11231.
Lawrence Livermore National Laboratory IM number: LLNL-JRNL-791193

\bibliography{reference_ml}

\end{document}